\documentclass[11pt,a4paper]{article}

\usepackage[margin=2.5cm]{geometry}
\usepackage{amsmath}
\usepackage{array}
\usepackage{graphicx}
\usepackage{float}
\usepackage{placeins}
\usepackage{endnotes}

\usepackage[round,authoryear]{natbib}
\usepackage{url}
\usepackage{hyperref}
\hypersetup{hidelinks}

\newcommand{\urlDocsMelodyFeatures}{\url{https://dmwhyatt.github.io/melody-features/feature_catalogue/}}
\newcommand{\urlRepoMelodyFeatures}{\url{https://github.com/dmwhyatt/melody-features}}
\newcommand{\urlRepoAnalysis}{\url{https://github.com/dmwhyatt/Style-Classification-Analysis}}
\graphicspath{{outputs/figures/}}

\title{Computational Features for Symbolic Melody Analysis}
\author{David M.\ Whyatt$^{1}$ and Peter M.\ C.\ Harrison$^{2}$\\[0.6em]
\normalsize
Centre for Music and Science, Faculty of Music, University of Cambridge, United Kingdom\\
\small $^{1}$\texttt{dmw56@cam.ac.uk}\quad $^{2}$\texttt{pmch2@cam.ac.uk}}
\date{}

\begin{document}
\maketitle

\begin{abstract}
This paper addresses the general problem of extracting music-theoretic and psychological features from symbolically encoded melodies. We review existing melodic feature extraction toolboxes, enumerate their features, and organise them into a common taxonomy. We then describe a new software library that provides implementations of all of these features in a straightforward Python package. We then demonstrate the combined feature set on the Essen Folksong Collection, using the dataset to produce a series of style classification models. These models help us answer key questions about the interpretability and dimensionality of the feature set. Our results show excellent classification accuracy using the full feature set, and promising performance for an eight-dimensional factor-analytic solution that improves the interpretability of the classifier. We distribute our new toolbox as an open-source Python package, \texttt{melody-features}, which can easily be used in various applications within music analysis, music psychology, and music information retrieval.
\end{abstract}


\section{Introduction}

Almost every musical culture across the world uses structures we might identify as `melodies': non-overlapping sequences of notes organised through time \citep{mehr_universality_2019, Ozaki2024-kf}. The analysis of melodies is therefore a key concern in musicology and music computing.

A minimal description of a melody might just list a series of pitches (e.g. C4, C4, G4, G4, A4, A4, G4) and an associated set of onset times (e.g. 0.0s, 0.75s, 1.5s, 2.25s, 3s, 3.75s, 4.5s). However, melodic analysis typically draws out a much broader range of features that draw out important underlying structures in the melody, and/or reflect various psychological aspects of melody perception. For example, a tonal analysis might identify the notes with particular scale degrees (tonic, tonic, dominant, dominant, submediant, submediant, dominant); alternatively, a contour analysis might instead represent the melody as a series of directional relations ($\rightarrow, \uparrow, \rightarrow, \uparrow, \rightarrow, \downarrow$).

The literature provides many useful formal approaches for deriving melodic features of music-theoretic and psychological interest. For example, \cite{lewin_intervals_2007} formulated a rigorous framework for the mathematical description and analysis of musical intervals and invariances; \cite{huron1996melodic} proposed a method for classifying the melodic contour of any conceivable melody, according to its first, last, and average pitches; \cite{parncutt_pulse_1994} introduced a perceptual model of pulse salience based on note durations. Research such as this has supported various interesting applications in diverse domains such as earworm modelling \citep{jakubowski_dissecting_2017}, tonality estimation \citep{chew_tonal_2007, krumhansl_key-finding_2001, parncutt_revision_1988, temperley_probabilistic_2008}, melodic expectation analysis \citep{krumhansl_music_1995, narmour_analysis_1990, pearce_construction_2005}, and cross-cultural music analysis \citep{raman_effects_2016}.

The diversity of available theoretical frameworks and corresponding melody features presents a challenge to current music computing researchers. One issue is that the literature lacks a consolidated account of existing features; instead, feature definitions are scattered over many historic books, articles, and technical reports. A second issue is that many features do not have publicly available software implementations, and the available implementations are often hard to use, and use an awkward variety of programming languages.

This paper addresses these problems. We present a comprehensive set of melodic features collected from the music computing literature and organised into a structured taxonomy. We also present a corresponding extensive software package (written in Python) that implements all of these features, making it straightforward to derive these features from arbitrary melody corpora.

\section{Existing Feature Sets}
Our first step in constructing the feature set was to identify all publicly available toolboxes for symbolic melody analysis. We are not aware of any feature-based toolboxes dedicated to melody analysis that operate on source audio at this time, and this remains a valuable future contribution to the literature. We restricted consideration to software that treats melodies as sequences of notes, where each note has just a single pitch; we did not consider software for pitch/melody transcription, and we did not consider software for analysing within-note ornamentations (e.g. trills, portamenti, etc.). We thereby found seven relevant toolboxes: FANTASTIC, SIMILE, IDyOM, jSymbolic, MIDI Toolbox, MUST, and Partitura. These toolboxes cover different parts of the music computing field, including both music cognition (FANTASTIC, IDyOM) and music information retrieval (jSymbolic, Partitura).

By the nature of the existing literature in this area, our perspective is necessarily Western-centric, often drawing on conventions from Western music theory. Though this limitation is unfortunate, it is often still possible to capture some cross-cultural aspects of music (e.g. using a corpus of representative music), and researchers ought to be open to using these approaches where relevant.

We will introduce each of the identified toolboxes in the sections that follow.

\subsection{FANTASTIC}
Feature ANalysis Technology Accessing STatistics (In a Corpus), shortened to FANTASTIC, was written in R \citep{mullensiefen_fantastic_2009}.\endnote{\url{https://www.doc.gold.ac.uk/isms/mmm/?page=Software\%20and\%20Documentation}} In FANTASTIC, melodies are symbolically represented using a combination of vectors corresponding to different melodic domains, such as a pitch sequence or onset times. This allows the user to compute a vast number of summary statistics across many different melodic domains. However, most of these summary statistics fail to preserve the temporal organisation of melodies, a property that is essential to understanding how melodies are perceived \citep{dowling_recognition_1972}. 

This problem motivates the construction of what are termed the m-types of a melody, which function as melodic \textit{n}-grams (discussed in Section~\ref{sec:mtypes}). Using this approach, it is possible to relate implicitly learned musical structures to prominent work on linguistic grammar and syntax \citep[e.g.][]{chomsky_syntactic_1957, lerdahl_generative_1996}.

\subsection{SIMILE and melsim}
SIMILE was developed in C++ to evaluate the similarity of two or more melodies \citep{mullensiefen_simile_2004}.\endnote{The source code for this software is available on request from the original authors.} Many features and representations relevant to producing summary statistics for a given melody are also relevant to calculating similarity, and as such, there is a large overlap between the design and implementation of SIMILE and FANTASTIC. However, SIMILE has not been maintained, and has become difficult to use.

Melsim serves as an effective update to SIMILE, currently under development by Sebastian Silas and Klaus Frieler in R \citep{silas_github_nodate}.\endnote{\url{https://github.com/sebsilas/melsim}} This software adopts a modular approach to computing melodic similarity, separating the calculation into two components: a transformation and a similarity measure. A transformation refers to the isolation of a melodic domain, and the similarity measure refers to the specific algorithm for which the transformation is to be compared across two or more melodies. This provides a great deal of flexibility as to how a melody may be analysed, allowing the user to combine a target domain with a similarity measure at their whim.

\subsection{IDyOM}
Information DYnamics Of Music (IDyOM) was written in Common Lisp \citep{pearce_construction_2005}.\endnote{\url{https://github.com/mtpearce/idyom}} IDyOM seeks to construct variable-order Markov models that represent the conditional probability distribution for a given musical event taking place.

The initial output from IDyOM is the predictive probability distribution for each note, expressed in terms of the negative log likelihood. These likelihood values are often then used to compute notewise information content and entropy. These measures correspond to levels of melodic surprise, where high negative log probability values, information content and entropy correspond to high levels of surprise or unpredictability. 

IDyOM also implements Narmour's Implication-Realisation hypotheses, which do not typically require any statistical learning. These features are easy to interpret and powerfully used alongside the likelihood values to facilitate the analysis of melodic expectancy.

\subsection{jSymbolic}
jSymbolic implements an impressive 1500 features, spanning melodic and extra-melodic domains such as instrumentation and texture \citep{mckay_jsymbolic_2006}.\endnote{\url{https://github.com/DDMAL/jSymbolic2}} Written in Java, its reported feature count is partially inflated by returning individual histogram bins and by exposing many variants of the same basic operation (e.g. separate features for the fraction of melodic thirds versus perfect fourths). Even so, the toolbox covers a wide range of distinct features.

Unlike most reviewed toolboxes, jSymbolic often exposes intermediate representations and further features derived from them. One distinctive example is a pitch class histogram ordered by the circle of fifths, from which the skewness and kurtosis are made available as features. 

A limitation of this toolbox, however, is weaker engagement with cognitive models. For instance, tonality is read from a MIDI major/minor event rather than estimated with a cognitively-relevant algorithm such as Krumhansl-Schmuckler \citep{krumhansl_cognitive_1990}. Incorporating such approaches would further enhance an already extensive toolbox.

\subsection{MIDI Toolbox}
The MIDI toolbox was written in MATLAB in 2004, and was last fully updated in 2016 \citep{eerola_midi_2004}.\endnote{\url{https://github.com/miditoolbox/1.1}} It provides a new representation of MIDI data in the form of a notematrix (NMAT), which expresses musical information using a $7 \times \textit{L}$ matrix, where \textit{L} is the number of notes in the melody. The seven columns correspond to set MIDI attributes, such as pitch number and onset time. 

This toolbox includes some functions similar to those found in jSymbolic, such as pitch class histograms (albeit without a circle of fifths ordering). However, the MIDI toolbox contains a more limited number of summary statistics that could be used for a computational analysis. Instead, the MIDI toolbox is perhaps best suited to visualising symbolic data.

\subsection{MUST}
This MATLAB toolbox draws inspiration from the MIDI toolbox, representing melodies in a reduced NMAT form that preserves only the pitch numbers, onset times and durations of notes \citep{clemente_set_2020}.\endnote{\url{https://github.com/compaes/MUST}}. MUST was devised to study the properties of musical stimuli, such as those one might find in behavioural experiments or computational studies. The features implemented in MUST correspond to four high-level categories: asymmetry, balance, jaggedness, and complexity. We include all of these features in the `complexity' category of our taxonomy. 

\subsection{Partitura}
Partitura is a Python package for representing and manipulating symbolic music data, with a particular focus on handling MIDI data from music performances \citep{cancino-chacon_partitura_2022}.\endnote{\url{https://github.com/CPJKU/partitura}} It also provides a small number of music analysis functions; we have incorporated all such functions that are applicable to melody analysis.

\subsection{Other toolboxes}
A handful of other notable toolboxes are not reviewed in detail or included in our package. We will briefly elucidate the reasons for their exclusion here.

Humdrum is an influential toolkit for \texttt{kern}-encoded symbolic music \citep[see also][]{HumdrumR}, but offers few melody-analysis features beyond those already covered above.\endnote{\url{https://github.com/humdrum-tools/humdrum-tools}}

music21 provides a rich Python representation of Western scores and convenient score iteration, but no built-in catalogue of melodic features.\endnote{\url{https://github.com/cuthbertlab/music21}}

Melfeature (Jazzomat/MeloSpy) is designed for analysing symbolically-encoded monophonic jazz solos \citep{frieler_introducing_2013}.\endnote{\url{https://jazzomat.hfm-weimar.de/download/download.html}} It is only partly documented \citep{PfleidererMartin2017ItJN}, not open-source, and many of its features overlap FANTASTIC and jSymbolic. Distinctive ideas such as compression/division complexity and self-similarity are complex tasks deserving of systematic treatment, so we leave them for future work.

\section{Theoretical Review}
\label{sec:review}
The reviewed toolboxes all share a common operating principle: they take one or more symbolically encoded melodies as an input, and compute various \textit{features} from those melodies. Each feature is intended to capture some meaningful aspect of the melody's structure, or of the way that structure is represented in the mind of the listener. These features can then be used as the input to statistical models, machine-learning models, and/or cognitive models.

We proceeded by compiling a long list of the features implemented by these toolboxes. We pruned this list by removing duplicate features as well as removing a small number of features with uncertain mathematical foundations (see Section~\ref{sec:excluded}); we also added a few features that seemed naturally motivated by the theoretical context.

The resulting list numbers 282 features (Supplementary File~1 Table~1). We have organised these features into 12 conceptual categories, each corresponding to a distinct aspect of melodic structure (e.g. interval, metre, tonality, complexity). We then have organised each category into one of three higher-level groups: pitch, rhythm, or pitch and rhythm combined.

The rest of this section now introduces these features in turn, following the conceptual categories outlined in Supplementary File~1 Table~1.

\subsection{Pitch}
\subsubsection{Absolute pitch}
A simple example of a computational representation of `absolute pitch' is MIDI key number. The MIDI key number may be calculated from an arbitrary frequency using the following equation:
\begin{equation}
\label{eq:midi_pitch}
n = \left\lfloor 69 + 12 \cdot \log_2\left(\frac{f}{440}\right) + 0.5 \right\rfloor
\end{equation}

\noindent Note that this particular expression assumes twelve-tone equal temperament with a reference pitch of A4 = 440 Hz.

Several descriptive features can then be computed from this absolute pitch representation. In particular, we define \textbf{\textit{Mean Pitch}} as the arithmetic mean of all the pitch values in a given melody; we then define \textbf{\textit{Pitch Standard Deviation}} as the sample standard deviation of these pitches, capturing their variability about the mean.

\subsubsection{Pitch interval}
Absolute pitch is not consciously accessible to most Western listeners; rather, melody perception is dominated by the appraisal of relative pitch relationships \citep{deutsch2013absolute, ward_8_1999}. These relationships can be expressed in terms of `intervals' between notes. Intervallic distance is perceived as logarithmic in frequency, or linear in the pitch scale described in Equation~\ref{eq:midi_pitch}. For example, for the notes C4 and G4, the pitch interval is seven semitones.

It is possible to express a melody as a series of pitch intervals. A common option is to express each note as an interval from the previous note, for example $[0, 7, 0, 2, 0, -2]$, but it is also possible to use different reference points, such as the first note in the melody or the tonic. Such intervallic representations reproduce the well-established cognitive principle of transpositional invariance, namely that a melody with the same pitch intervals is recognised as being essentially the same melody, regardless of the starting pitch \citep[see][]{chrisman_describing_1977, han_melody_2024, lattner_learning_2019}.

Pitch intervals are often summarised in the same way as absolute pitch values, producing measures such as \textbf{\textit{Mean Absolute Interval}} and \textbf{\textit{Pitch Interval Standard Deviation}}.

\subsubsection{Pitch contour}
The contour of a pitch sequence can be considered as a simplified encoding of the same information obtained from the pitch interval. However, instead of preserving the exact size of the difference between consecutive pitches, contour representations only preserve the approximate shape of the melody, privileging (changes of) direction over exact intervals. Changes in melodic contour are highly noticeable, and have even been demonstrated to be perceived pre-attentively under experimental conditions \citep{mittag_transitional_2016}. The cognitive encoding of a contour is less complex than a series of pitch intervals, and as a result is more readily stored in memory for novel melodies \citep{dowling_time_1995}.

Pitch contours may be represented in a number of different ways. One example is an \textbf{\textit{Interpolation Contour}}, produced by linearly interpolating between turning points in the pitch sequence \citep{mullensiefen_fantastic_2009}. Another example is \textbf{\textit{Parsons' Contour}}, which simply codifies the direction of each interval, discarding the size of the interval altogether \citep{parsons_directory_1975}. One further example is \textbf{\textit{Huron Contour}}, which reduces the pitch information to simply the size of the interval between the start pitch and mean pitch, and the mean pitch to final pitch \citep{huron1996melodic}, and then classifies the resulting shape into one of nine categories. However, recent work has called into dispute whether melodic contours can indeed be classified into discrete categories based on their apparent similarity \citep{cornelissen2026melodiccontourdoescluster}.

\subsubsection{Pitch class}
A pitch class can be defined as an equivalence class over an octave transposition. The representation therefore captures the widespread perceptual phenomenon of \textit{octave equivalence}, whereby pitches separated by a whole number of octaves are perceived as highly similar \citep{stumpf_origins_2012}. Pitch classes are mostly used when the counts of different notes is important, but not their pitch height, such as in tasks involving key estimation. Pitch classes are often described using similar measures to pitches and pitch intervals, or used to measure the distribution of notes across a given scale. For example, \textbf{\textit{pcdist1}} describes the distribution of pitch classes across the 12-tone chromatic scale \citep{eerola_midi_2004}.

\subsubsection{Tonality}
Tonality refers to the key of a melody or piece of music. Although this is often made evident by harmonic context (e.g. chordal accompaniment), Western listeners can often infer key from the pitches of a melody alone. This motivates the need to compute the implicit tonality of a melody from its pitch content.

The standard approach to this task uses the Krumhansl-Schmuckler algorithm \citep{krumhansl_cognitive_1990}. This algorithm uses major and minor key profiles, derived from empirical human rating data of how well notes `belong' to each key \citep{krumhansl_tracing_1982}. For an arbitrary melody, pitches are aggregated into a distribution of pitch classes and this distribution is compared with each of the 24 possible key profiles using Pearson's correlation; the highest correlation is selected as the implied key.

Previous findings, such as a working-memory advantage for strongly tonal melodies \citep{schulze_working_2012}, motivate features capable of estimating perceived tonal strength. For example, \textbf{\textit{Tonal Clarity}}, the ratio of the two highest profile correlations, measures how distinctly listeners are likely to perceive the implied tonic \citep{mullensiefen_fantastic_2009}. Similarly, it has been observed that listeners may also categorise pitches as scale degrees once a key is inferred, making out-of-key notes highly noticeable \citep{raman_effects_2016}; \textbf{\textit{In Scale}} marks, for each note, whether it belongs to the implied tonality.

An alternative, implemented in Partitura and the present package as \textbf{\textit{Tonal Tension}}, is Chew's Spiral Array \citep{chew_tonal_2007}: a three-dimensional pitch spiral obtained by expanding the \textit{Tonnetz} \citep[see][]{cohn_neo-riemannian_1997}. This model does not assume octave invariance, and representing pitches instead of pitch classes may improve polyphonic key estimation relative to Krumhansl-Schmuckler \citep{chuan_polyphonic_2005}. The model is geometric, however, so its perceptual grounding is less direct than rating-based key profiles. The Krumhansl-Schmuckler model is therefore generally preferred in music cognition.

\subsection{Rhythm}
\subsubsection{Timing}
The timing of a musical note is captured by its `onset' and `offset' times, corresponding to the beginning and end of the note respectively. These times may be expressed in units of time (e.g. seconds, milliseconds), or score units (e.g. quarter notes). The duration of a note is then calculated as the difference between its onset and offset time.

\subsubsection{Inter-onset interval}
When multiple notes are played in sequence, forming a rhythmic pattern, it is common to compute inter-onset intervals (\textbf{\textit{IOI}}), defined as the difference between successive onset times (c.f. analogous to pitch intervals, defined as the difference between successive pitches). As above, inter-onset intervals may be expressed in either temporal units or score units. For example, a rendition of the opening rhythm to Happy Birthday could be written as [0.75, 0.25, 1, 1, 1].

A given rhythmic pattern may be performed faster or slower without disrupting its cognitive identity. To capture this concept of tempo invariance, one can compute inter-onset ratios (\textbf{\textit{IOI Ratio}}) by dividing each inter-onset interval by the inter-onset interval that immediately precedes it. For Happy Birthday, the above rhythm could be written as [0.333, 4.0, 1.0, 1.0].

\subsubsection{Metre}
Metre refers to the repeating temporal grid on which musical events are organised. A minimal model uses two levels: pulses per cycle and the baseline pulse duration (\textbf{\textit{Meter Numerator}} and \textbf{\textit{Meter Denominator}}). Western music most often uses four quarter-note pulses per cycle. Richer metre models add further levels down to \textit{tatums}, the smallest beat subdivision present in a piece \citep{a5894967f2324f9cb0aada8e3e7c9a22}.

Note placement on the metric grid can produce syncopation by emphasising weak beats once a regular metre is established \citep{sioros_syncopation_2014}. One such syncopation model provides the definition of \textbf{\textit{Syncopation}} as the result of a note on a metrically weak beat followed by a rest on the subsequent strong beat \citep{longuet-higgins_rhythmic_1984}. Notes are stratified by duration into a metrical hierarchy, assigned weights (strong beats larger than weak), and syncopation is the difference in weight between that weak beat and the following strong-beat rest. This model influenced later syncopation models, including \cite{witek_syncopation_2014, hoesl_modelling_2018}. As a rule-based model, however, this cannot capture micro-timing or tempo effects. Because syncopation adds rhythmic complexity and can make melodies harder to encode in memory \citep{fitch_perception_2007}, such features are also relevant to music cognition.

\textbf{\textit{Syncopicity}} offers a complementary definition \citep{mullensiefen_simile_2004}: a syncope occurs when a note misses the grid at one metrical level, falls on the grid at the level below, and has an IOI longer than that lower level. Syncopicity sums  the number of syncopes divided by the number of notes, over all levels. Relative to the Longuet-Higgins and Lee model, it aggregates overlapping metrical levels into one melody-level score, whereas LHL yields note-wise syncopation values.

Most reviewed toolboxes also assume a fixed metre. However, metre can change by altering pulses per cycle, the baseline pulse, or both. We leave the task of fully modelling metre change to future work, but we introduce \textbf{\textit{Proportion of Time in First Meter}} as a simple index of metre change within a melody.

\subsection{Pitch and rhythm}
\subsubsection{Lexical diversity}
\label{sec:mtypes}
Whilst pitches, onsets, and their derivatives can all be meaningfully analysed in isolation, it is also possible to analyse melody using an approach that combines them into a unified representation. One combined representation is defined by \cite{mullensiefen_fantastic_2009} as the m-types of a melody, which are constructed as follows:

First, the melody is segmented into phrases, which are identified by a gap between notes of a specified length. Next, a window of length \textit{n} notes is moved over each phrase, capturing the sequence of notes in order. Then, the pitch contents are transformed to pitch intervals, and classified according to their direction and magnitude. Similarly, the onsets are transformed to IOIs, which then are transformed to IOI ratios. These ratios are classified into `shorter,' `equal,' or `longer' to capture the relationships between consecutive notes in a tempo-invariant form. These pitch interval classes and IOI ratio classes are paired together to produce an `m-token,' representing the combined pitch and rhythm relationships of the adjacent \textit{n} notes in the melody. This process is repeated for every note position in the melody for which a window of size \textit{n} is able to be constructed. The total collection of m-tokens of all lengths is referred to as the m-types of a melody. 

M-types capture the relationship between pitch and rhythm explicitly under one representation. This allows powerful statistics to be computed for the captured melodic movements relevant to repetition or diversity in the m-tokens. These statistics provide an easy interface for relating musical analysis to linguistic studies on concepts such as vocabulary. An example of these features is \textbf{\textit{Simpson's D}}, which captures the amount of repetition in the m-tokens.

\subsubsection{Complexity}
Cognitively, complexity is the processing demand a stimulus places on a listener. Stimuli that are more informative, less familiar, or less predictable are harder to process. Algorithmic measures of complexity can be effective, but vary in their cognitive relevance. For example, Kolmogorov complexity, the length of the shortest program that regenerates a sequence \citep{Kolmogorov01011968}, can capture pattern encoding, but is not itself a good cognitive model because the program need not use operations natural to human minds. Most musical complexity algorithms therefore take a statistical approach, drawing unpredictability metrics from information theory and vocabulary diversity measures from computational linguistics.

One clear difference between algorithmic approaches concerns whether a measure accounts for a reference corpus, which is thought to inform the listener's judgements. For instance, a melody may be considered highly unpredictable and complex by listeners unfamiliar with the musical idiom that informs it, but relatively predictable and simple by a different group of listeners who recognise the idiom and can therefore compress their mental representation of the melody. 

The simplest melodic complexity algorithms do not rely on a reference corpus. Instead, they count occurrences of each value of a discrete attribute (e.g.\ pitch, pitch interval, pitch class, IOI), treat these counts as a probability distribution, and compute the Shannon entropy of the distribution \citep{shannon_mathematical_1948}. This yields features such as \textbf{\textit{Pitch Entropy}}, \textbf{\textit{Interval Entropy}}, and \textbf{\textit{Duration Entropy}} \citep{mullensiefen_fantastic_2009}. Various other measures can be constructed in a similar way, replacing entropy with measures from computational linguistics for quantifying the diversity of distributions. For example, \textbf{\textit{Yule's K}} measures the rate of repetition for tokens in a sequence -- other measures include \textbf{\textit{Sichel's S}} and \textbf{\textit{Honore's H}}. 

\subsubsection{Expectation}
Musical expectation may be produced and shaped by any one of the reviewed domains. It is often experienced as surprise when a melody violates a listener's developing mental schema, for example where a note violates the inferred tonality of a melody.

One influential expectation model is Narmour's Implication-Realisation (IR) framework \citep{narmour_analysis_1990}, which uses pitch intervals. Though it includes a top-down component reflecting a listener's musical exposure, expectancy can also be formalised from the bottom-up without that exposure \citep{krumhansl_music_1995}. The bottom-up component comprises five principles relating the final two intervals of a sequence (the implicative and realised intervals). For example, \textbf{\textit{Narmour Registral Direction}} asserts that (a)~a large implicative interval ought to be followed by a change in direction and (b)~a small implicative interval ought to continue in the same direction; a melody scores~1 if either criterion is met and~0 otherwise \citep{krumhansl_music_1995}. Later work found redundancy among the five principles and comparable predictive power from a reduced subset \citep{schellenberg_expectancy_1996, schellenberg_simplifying_1997}.

One criticism of the IR outputs is that they are discrete conformity scores and do not quantify predictive uncertainty. Probabilistic models address this issue. \cite{conklin_multiple_1995} used multiple-viewpoint Markov models to produce note-wise entropy profiles, and thus estimates of surprise that can be averaged over a melody, drawing on both long-term (LTM; corpus-trained) and short-term (STM; within-melody) memory. IDyOM developed this approach further with variable-order Markov models, returning note likelihoods (and derived information content and entropy) and combining LTM and STM predictions using weighted geometric mean \citep{pearce_construction_2005}.

An alternative probabilistic approach estimates the implied key and likelihood for melodic continuations using Bayesian reasoning \citep{temperley_probabilistic_2008}. This method performs well for pitch continuation, key finding, and wrong-note detection, but does not account for rhythm, which strongly shapes expectancy \citep{schmuckler_performance_1990, steedman_perception_1977}.

\subsubsection{Corpus}
Lexical diversity, complexity, and expectation modelling may also involve comparison to a corpus. Within tasks involving lexical diversity, it is common to model how rare a token is by comparison to a reference collection of tokens. One such approach involves computing `document frequencies,' which count the number of items in a corpus that contain a given token. From these document frequencies, a simple measure of token rarity is \textbf{\textit{Mean Document Frequency}}, which corresponds to the average rarity of tokens in a melody. This approach could theoretically be applied to any of the reviewed melodic domains, but is most commonly applied to m-types.

Similarly, complexity and expectation models may be improved by accounting for familiarity with a given musical style: if a listener has a high degree of exposure to highly intervallic and unpredictable melodies, then their perceived level of complexity may be lower than that of a listener who is unfamiliar with this style. This can be partially controlled for by supplying a corpus that simulates the listener's musical exposure.

\section{Features for Symbolic Melody Analysis}
Now that we have identified and organised the available sources of computational features for melody analysis, we may discuss the details of those we believe to be relevant. The table that follows is produced from a reduction of the features found in the reviewed toolboxes: we retain only the features most pertinent to a cognitive analysis of melody with minimal redundancy.

Supplementary File~1 Table~1 details the source of each feature and briefly describes how it is produced. An interactive version of this table is available in the package documentation, with hyperlinks that direct to the source code implementation.\endnote{\urlDocsMelodyFeatures}

Some of the reviewed features return statistics that explicitly describe the contents of a melody on a note by note basis: for instance, IDyOM's negative log likelihood scores can be computed for every note in the melodic sequence. As such, the table also includes a column that identifies whether a feature produces a sequence of values, or in some way describes the original representation with a single value or distribution.

Where the `pre-existing implementations' column is populated with `novel', these features are newly proposed in line with existing features found in the literature. For example, the reviewed works do not include \textbf{\textit{Mean Duration}} as a measure of average note duration in quarter-note units, but this can easily be introduced to produce a descriptor analogous to the other quarter-note-based rhythm features. These features seek to expand melodic domains that appear to be under-represented in the review table.

\section{Implementation}
The present Python package contains all of the features catalogued in Supplementary File~1.\endnote{\urlRepoMelodyFeatures} The majority of features found in the table are newly implemented in Python. \textit{melsim} is included with the feature set using a wrapper approach, allowing the user to interface with the R package through a Python front-end. However, calculating the similarity between melodies is a more specific use case than that of the rest of the feature set, so the user must invoke Melsim separately to the rest of the feature calculations.

The package is released under the MIT licence. Code adapted from Partitura is licenced under the Apache Licence~2.0, and the Essen Folksong Collection distributed with the package is licenced under CC BY-SA~4.0 \citep{eck_essen_2024, schaffrath_essen_1995}. 

\subsection{Definition changes}
Our final feature set differs in some important ways from the contributing toolboxes. These differences are documented in the Notes column of Supplementary File~1 Table~1. Some differences correspond to bugs in the reference implementations that we have fixed in our implementation (e.g. \textbf{\textit{Minor Major Third Ratio}}, \textbf{\textit{Interpolation Contour}}, \textbf{\textit{Step Contour}}). Some correspond to features that are duplicated across toolboxes, in which case we only retain one version.

\subsection{Excluded features}
\label{sec:excluded}
We excluded a small number of toolbox features for the following reasons. We omitted nine variants from FANTASTIC  (Features 46-54) that substitute m-type corpus frequencies or TF-IDFs into established diversity/entropy formulae. These substitutions lack a clear theoretical rationale and were often hard to interpret in practice. We retained all other named and numbered features from the FANTASTIC technical report.

We also omitted features from jSymbolic that require polyphonic input (e.g. per-voice attack timing and partial-rest statistics), features based on pitch-bend MIDI events (glissando, vibrato, and microtone measures), and \textit{Major or Minor}, which essentially duplicates FANTASTIC's \textit{Mode} feature.

We considered most of the measures implemented in SIMILE to be included in \textit{melsim}, which we cover in entirety by wrapping the \textit{R} library. Only a handful of features from SIMILE are retained by the present package as a result. 

\subsection{Style classification analysis}
Our feature set can be applied to a wide range of tasks. Here, we demonstrate one such application by building a series of style classification models that attempt to predict the region of origin for an approximately balanced set of melodies from Europe and China, similar to the task presented in \cite{KLARLUND2023105405}.

All scripts used to run the following analyses and produce the following figures are made available online on GitHub.\endnote{\urlRepoAnalysis}

\subsubsection{Data}
The melodies used in this analysis were gathered from Eck's conversion and tokenisation of the Essen Folksong Collection \citep{eck_essen_2024, schaffrath_essen_1995}, which we include in the Python package.\endnote{\url{https://www.kaggle.com/datasets/sebastianeck/essen-folksong-database-conversion-and-tokenization}} This dataset contains 8,472 unique melodies. Two of the MIDI files contained corrupted key signature information and therefore could not be read, and a further 91 melodies were detected as containing overlapping durations, so were excluded from the feature calculation process. Our dataset numbers 8,379 melodies as a result.

We produced a balanced subset of melodies from two large geographical regions represented in the Essen Folksong Collection: Europe and China. This collection contains 2,173 monophonic melodies that are identified as Chinese,\endnote{\url{https://kern.humdrum.org/cgi-bin/browse?l=essen/asia/china}} and 6,212 melodies identified as European.\endnote{\url{https://kern.humdrum.org/cgi-bin/browse?l=/essen/europa}} We randomly sampled from these 6,212 European melodies to produce a subset of 2,192 European melodies, yielding an approximately balanced dataset of 4,365 melodies ($50.22\%$ European).

All 282 features implemented in the package were computed for each of these 4,365 melodies. We dropped any non-numeric scalar features and standardised the remaining features to zero-mean before using these features to train the model. Zero-variance features after standardisation were also dropped. Features that are computed relative to a corpus (FANTASTIC \& IDyOM LTM models) were trained on the 903 Western traditional melodies utilised in \cite{pearce_statistical_2018}. Any melodies that appear in this corpus were excluded from our Essen-European subset to prevent leakage during model training. A total of 235 numeric features were used to produce each of the models that follow.

\subsubsection{Logistic regression}
\begin{figure}[!htbp]
  \centering
  \includegraphics[width=\textwidth]{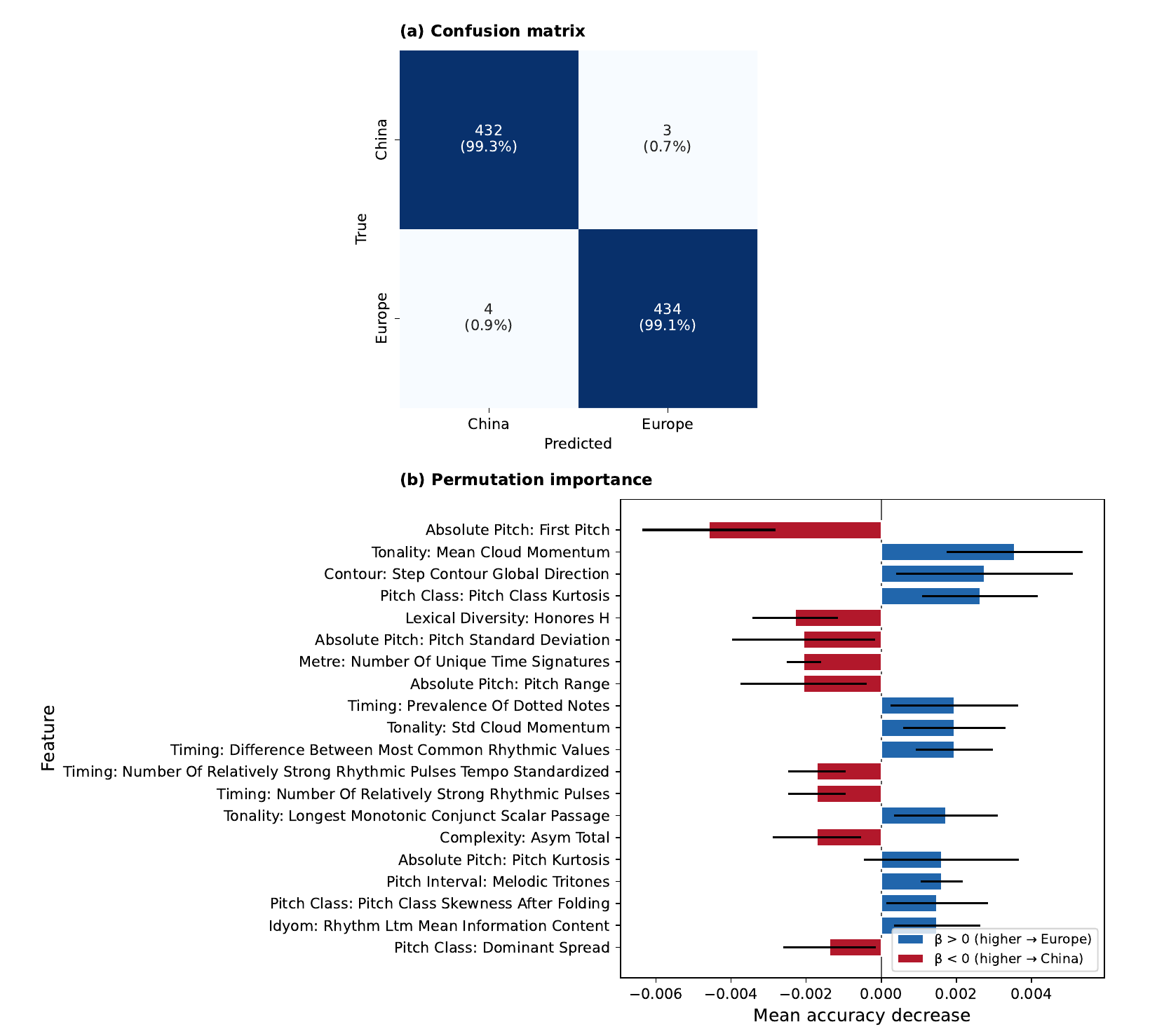}
  \caption{Logistic regression style classification: (a)~confusion matrix on the held-out test set; (b)~permuted feature importance (signed by logistic coefficient).}
  \label{fig:logreg}
\end{figure}

We performed a logistic regression on a training set of 3,492 melodies, which is stratified to preserve the approximate balance of the full dataset. This training set represents $80\%$ of the total dataset. The remaining 873 melodies were withheld for testing the final model performance. Over 5-fold stratified cross-validation, the model produced a mean accuracy of $0.9874 \pm 0.0029$. On the test set, the model achieved an accuracy of $0.9920$, misclassifying seven melodies across the two possible geographies (Figure~\ref{fig:logreg}a).

\FloatBarrier

To better understand how this classification model was making its predictions, we tested the permuted feature importance for each of the top level groups presented in the earlier Section~\ref{sec:review}: Pitch, Rhythm, and Pitch \& Rhythm. We follow the methodology of \cite{breiman_random_2001} by randomly shuffling the values of all features in the test set for a given taxonomy group. We then measure the mean decrease in classification accuracy across each permutation to capture the importance of that specific feature group, and repeat this method across all three groups. Each feature group was randomly permuted 10 times.

Of the three groups, Pitch features contributed most to the classifier (mean importance~$0.4345$). Rhythm features had a modest contribution (M.I.~$0.05762$), but Pitch \& Rhythm features (which includes Corpus features) were not very useful to the classifier (M.I.~$0.002405$). We may conclude from this result that pitch-based features, which encompass intervals, contours, and tonality, offer the most valuable information for classifying melodies as either Chinese or European.

We can deepen our understanding of how the classification model works by measuring the permuted feature importance for each of the features implemented in the package.
Each feature was randomly permuted 10 times. Chinese melodies were most positively predicted by variability in rhythmic events and wide pitch range, whereas European melodies were most positively predicted by features related to Western tonal concepts (Figure~\ref{fig:logreg}b).

\subsubsection{Exploratory factor analysis}
Many of the 235 features are highly correlated, which under certain circumstances can make results difficult to interpret. We sought to identify a low-dimensional representation that remains useful for the classification task while using a smaller set of music-theoretic constructs that are interpretable and suitable for smaller datasets.

We attempted a Principal Component Analysis (PCA), but found that the components this yielded were difficult to interpret, perhaps because the requirement of orthogonal components is inconsistent with the latent structure of musical features. Instead, we used Exploratory Factor Analysis (EFA) with the oblique `promax' rotation, which does not impose orthogonality. We then used the scores obtained from these factors as covariates in another logistic regression model.

\begin{figure}[!htbp]
  \centering
  \includegraphics[width=0.85\columnwidth]{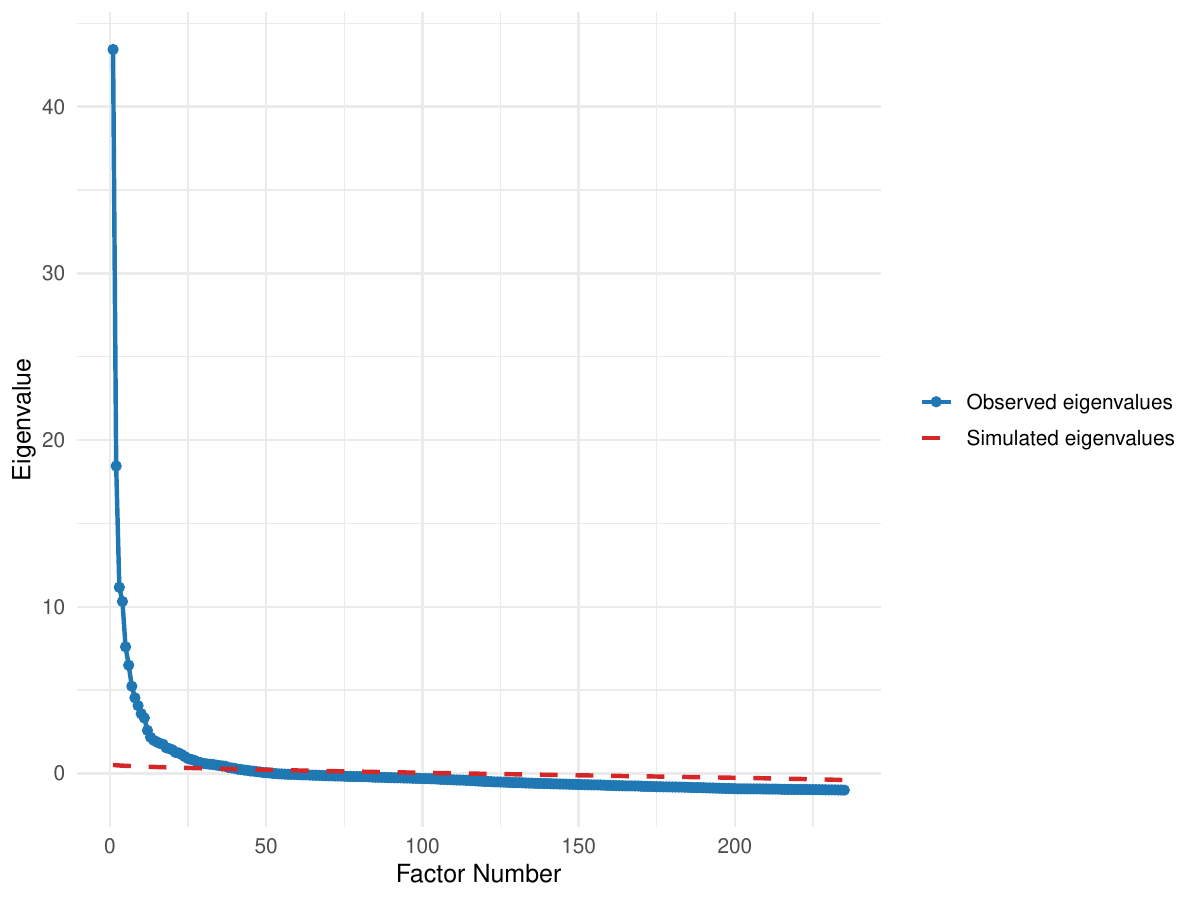}
  \caption{Scree plot of eigenvalues for 235 factors.}
  \label{fig:scree}
\end{figure}

Inspection of the scree plot of eigenvalues reveals a clear elbow after the eighth factor (Figure~\ref{fig:scree}). As such, we extracted the first eight factors, which explain $46.27\%$ of the total variation in melodic features (Table~\ref{tab:efa-variance}). These eight factors were then interpreted by inspecting the strongest loadings on each factor, e.g. Factor 1 has heavy loadings for features corresponding to longer average note durations (\textbf{\textit{Mean Rhythmic Value, Average Note Duration, Mean Duration}}), so the label `Long Rhythms' appears appropriate. The top ten feature loadings for each factor are included in Supplementary File~1 Table~2. We include a 3D visualisation of this factor solution with the reproduction scripts for this analysis.

\begin{table}[!htbp]
  \centering
  \begin{tabular}{@{}clcc@{}}
    \hline
    Factor & Interpretation & Prop.\ Var.\ (\%) & Cum.\ Var.\ (\%) \\
    \hline
    1 & Long Rhythms & 15.75 & 15.75 \\
    2 & Irregular Rhythms & 5.56 & 21.30 \\
    3 & Pitch-Class Variety & 4.53 & 25.84 \\
    4 & Overall Complexity & 4.45 & 30.29 \\
    5 & Wide Intervals & 4.37 & 34.65 \\
    6 & Dense Rhythms & 4.21 & 38.87 \\
    7 & Stepwise Complexity & 4.13 & 43.00 \\
    8 & Corpus Familiarity & 3.27 & 46.27 \\
    \hline
  \end{tabular}
  \caption{EFA: proportion and cumulative variance explained by each factor.}
  \label{tab:efa-variance}
\end{table}
\FloatBarrier

\begin{figure}[H]
  \centering
  \includegraphics[width=\textwidth]{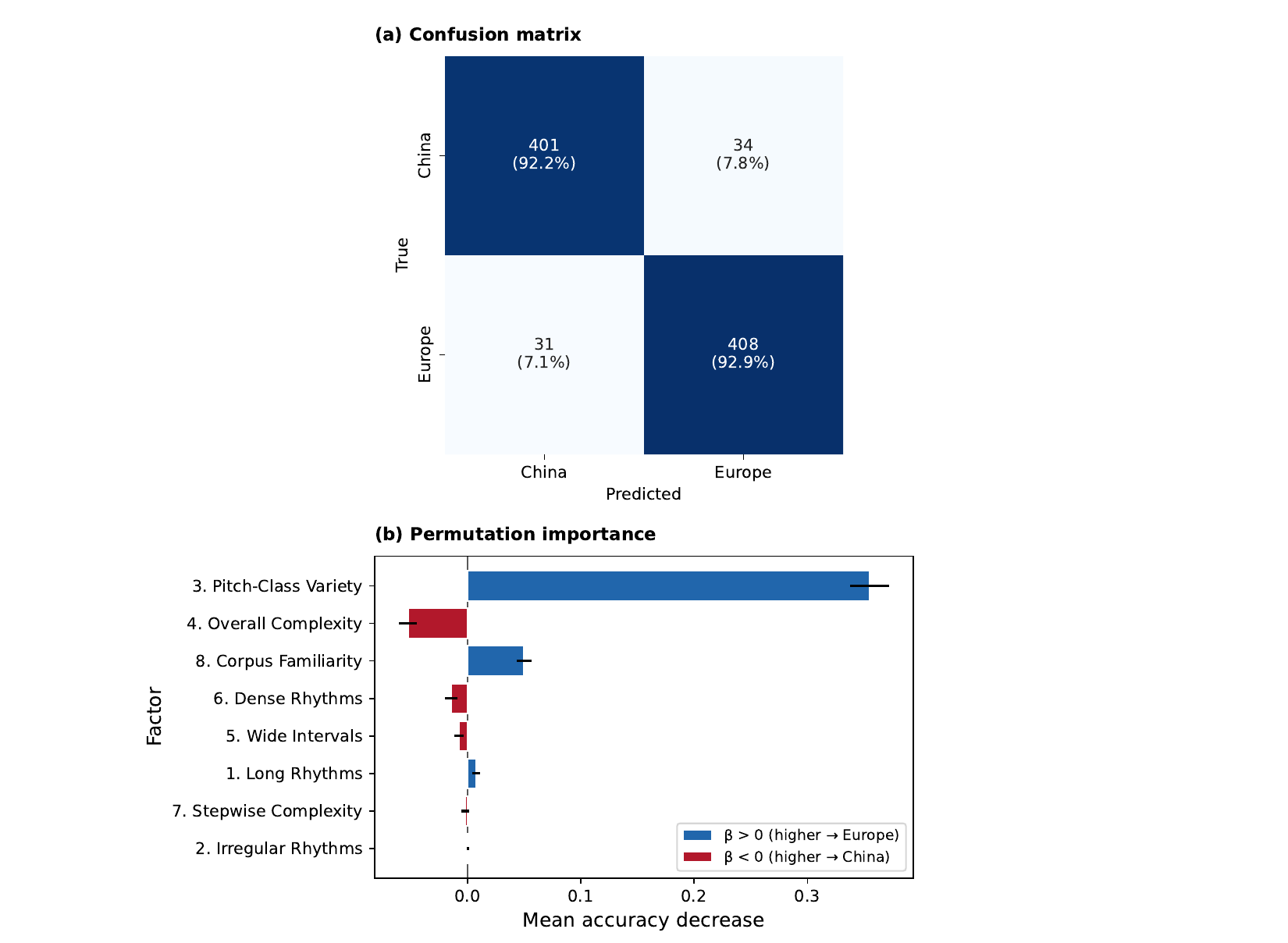}
  \caption{Factor-analytic logistic regression: (a)~confusion matrix on the held-out test set; (b)~permuted factor importance (signed by logistic coefficient).}
  \label{fig:efa}
\end{figure}

Factor scores were computed for each of the eight factors, and a logistic regression was trained using these scores. 5-fold cross-validated accuracy for this model was $0.9433 \pm 0.0038$, and the final test set accuracy was $0.9256$ (Figure~\ref{fig:efa}a).
Permutation testing revealed that complex melodies with dense rhythms and wide pitch intervals were more likely to be Chinese in origin, whereas melodies with a wider variety of pitch classes that were more similar to the reference corpus were more likely to be European (Figure~\ref{fig:efa}b).
Relative to the full feature set, these eight factors greatly reduce dimensionality. Approximately $7\%$ of classification accuracy is lost using this solution, making it a viable option for applications where datasets are small or robust interpretability is essential.

\subsubsection{Comparison across toolboxes}
As a final step, we quantitatively compared the predictive performance of the different toolboxes that we have reimplemented here. We stratified the numeric features calculated earlier into categories corresponding to their original implementation and trained separate logistic regression models using only features corresponding to each original implementation. As before, we performed 5-fold cross validation (Table~\ref{tab:source-comparison}).

Performance varies across the toolboxes, but the strongest approach ceiling accuracy. As might be expected from our consolidatory approach, our combined toolbox obtains the highest accuracy. Future work could probe performance with more challenging tasks, for example distinguishing musics of different European countries from another.

\begin{table}[!htbp]
  \centering
  \begin{tabular}{@{}lccccc@{}}
    \hline
    Source & No. features & CV acc.\ & CV SD & Train acc.\ & Test acc.\ \\
    \hline
    \textit{\textbf{melody-features}} & \textbf{235} & \textbf{0.9874} & \textbf{0.0029} & \textbf{0.9991} & \textbf{0.9920} \\
    jSymbolic & 125 & 0.9825 & 0.0048 & 0.9931 & 0.9908 \\
    FANTASTIC & 40 & 0.9696 & 0.0045 & 0.9725 & 0.9794 \\
    MUST & 20 & 0.9562 & 0.0098 & 0.9585 & 0.9668 \\
    Novel & 18 & 0.9313 & 0.0034 & 0.9330 & 0.9381 \\
    MIDI Toolbox & 22 & 0.9307 & 0.0087 & 0.9350 & 0.9427 \\
    Partitura & 6 & 0.8634 & 0.0108 & 0.8634 & 0.8740 \\
    IDyOM & 18 & 0.8577 & 0.0086 & 0.8631 & 0.8625 \\
    SIMILE & 3 & 0.7649 & 0.0144 & 0.7660 & 0.7973 \\
    \hline
  \end{tabular}
  \caption{Logistic regression by feature extraction source.}
  \label{tab:source-comparison}
\end{table}
\FloatBarrier

\section{Discussion}
This paper consolidates the existing computational approaches to melody analysis into a shared taxonomy, and releases this consolidation as an open-source Python package to support future research. This package is intended as an easy entry point for melodic features that have previously been scattered across various different software resources and research papers.

Our style classification task, which uses both the full feature set and a lower-dimensional EFA solution, demonstrates a practical application of the package. The near-ceiling performance of the full feature set shows strong potential for MIR tasks, whereas the comparable performance of the eight-factor solution demonstrates a more interpretable approach that may be suitable for music psychology research.

Despite its novelty, our package is limited in many similar ways to those reviewed. Presently, we cannot robustly process polyphonic melodies, so we prevent the package from attempting to compute features for any melodies detected as polyphonic. Using a source-separation algorithm to isolate monophonic lines from a polyphonic melody would enable the operation of our package, but we do not attempt to implement such an algorithm at this time.

Our package is also limited with regards to features corresponding to tonality, due to their reliance on the Krumhansl-Schmuckler key-finding algorithm. A major limitation of this algorithm is that its templates only cover two modes (major and minor), limiting its stylistic generalisability. A valuable future project could be to extend the algorithm to function for arbitrary musical scales, following the example of \cite{huron_cognitive_2006}.

\FloatBarrier
\section*{Notes}
\theendnotes

\section*{Reproducibility}
The following resources are publicly available:
\begin{itemize}
  \item Package: \urlRepoMelodyFeatures
  \item Documentation and feature catalogue: \urlDocsMelodyFeatures
  \item Style-classification analysis scripts: \urlRepoAnalysis
\end{itemize}

\section*{Acknowledgements}
The authors would like to thank Klaus Frieler for his useful insights during the early stages of this work.

\bibliographystyle{apaTISMIR}
\bibliography{references}
\end{document}